\documentstyle[12pt,epsf]{article}

\newcommand{\beq}{\begin{equation}}
\newcommand{\eeq}{\end{equation}}
\newcommand{\bea}{\begin{eqnarray}}
\newcommand{\eea}{\end{eqnarray}}
\newcommand{\nn}{\nonumber}
\newcommand{\ra}{\rightarrow}

\newcommand{\gtrsim}{\ \rlap{\raise 2pt\hbox{$>$}}{\lower 2pt \hbox{$\sim$}}\ }
\newcommand{\lessim}{\ \rlap{\raise 2pt\hbox{$<$}}{\lower 2pt \hbox{$\sim$}}\ }

\newcommand{\np}[1]{Nucl. Phys. {\bf #1}}
\newcommand{\pl}[1]{Phys. Lett. {\bf #1}}
\newcommand{\pr}[1]{Phys. Rev. {\bf #1}}
\newcommand{\prl}[1]{Phys. Rev. Lett. {\bf #1}}

\newcommand{\ijmp}[1]{Int. Jour. Mod. Phys. {\bf #1}}
 
\newcommand{\ptp}[1]{Prog. Theor. Phys. {\bf #1}}

\begin{document}

\thispagestyle{empty}
\null

\hfill CERN-TH-97-238

\hfill FTUV/97-53,IFIC/97-54

\hfill hep-ph/9709369

\vskip 1cm

\begin{center}
{\Large \bf      
How sensitive to FCNC can $B^0$ CP asymmetries be? 
\par} \vskip 2.em
{\large         
{\sc G. Barenboim$^1$, F.J.Botella$^1$, G.C.Branco$^2$ and O.Vives$^1$
}  \\[1ex] 
{1)Departament de F\'\i sica Te\`orica, Universitat 
de Val\`encia  and  IFIC, Centre Mixte Universitat 
de Val\`encia - CSIC} \\
{\it E-46100 Burjassot, Valencia, Spain} \\[1ex]
\vskip 0.2em
2) Theory Division, CERN, CH-1211 Geneva 23, Switzerland and 
CFIF/IST and Dept. Fisica, Instituto Superior Tecnico, Av. Rovisco Pais \\
{\it 1096 Lisboa, Codex Portugal}
\vskip 0.3 em
\par} 
\end{center} \par
\vfil
{\bf Abstract} \par
We show that the study of CP asymmetries in neutral $B$-meson
decays provides a very sensitive probe of flavour-changing neutral
currents (FCNC). We introduce two new angles, $\alpha_{SM}$ and
$\beta_{SM}$, whose main feature is that they can be readily obtained 
from the measurement of the CP asymmetries $a_{J/\psi K_s}$, 
$a_{\pi^+ \pi^-}$ and the ratio $R_u \equiv|V_{ud}V_{ub}^*|/|V_{cd}V_{cb}^*|$,
providing a quantitative test of the
presence of new physics in a model-independent way.

Assuming that new physics is due to the presence of an isosinglet
down-type quark, we indicate how to reconstruct the unitarity 
quadrangles and point out that the measurements of the above 
asymmetries, within the expected experimental errors, may 
detect FCNC effects, even for values of 
$\left| \sum_{i=1}^3 V_{id}
V_{ib}^* / (V_{td} V_{tb}^*)\right| $
 at the level of a few times $10^{-2}$.
\par
\setcounter{page}{0}
\clearpage

\section{Introduction}

The forthcoming experiments at $B$-factories will provide crucial tests of the
Standard Model (SM) and its Cabibbo--Kobayashi--Maskawa mechanism (CKM) for
flavour mixing and CP violation. The fact that in gauge theories CP
violation and flavour mixing arise precisely from their two least known
sectors, namely the Yukawa coupling and/or the Higgs sector, enhances the
importance of the future experiments on $B$-mesons.

At the moment, there is a considerable amount of data on the CKM mixing 
matrix, leading to the measurements of $\left|V_{ud}\right| $,
$\left| V_{us}\right| $, $\left| V_{ub}\right| $, $\left| V_{cb}\right| $, 
$\epsilon $, $\epsilon ^{\prime }$, $x_{d}$, etc. Since, for three
generations, the quark mixing matrix is completely fixed by four
parameters, the present experimental data lead in principle to an 
overdetermination of the CKM matrix. In practice, the situation is more 
involved, due both to experimental errors and to various hadronic 
uncertainties in extracting the values of $|V_{ij}|$ from the experimental 
data. The crucial role played by CP asymmetries in neutral $B$-meson decays
such as $B_d^0 \ra J/\psi K_S$, $B_d^0 \ra \pi^+ \pi^-$
stems from the fact that, being dominated by one weak decay amplitude,
they are free from most of the hadronic uncertainties
\cite{nir2}. In the SM, these CP 
asymmetries are given by
\bea
\label{cpasim}
a_{J/\psi K_S} & = &
- \frac{\Gamma\left(B^0 \longrightarrow J/\psi \, K_s\right)
 - \Gamma\left(\overline{B}^0 \longrightarrow J/\psi \, K_s\right)}{
 \left(\sin(\Delta M t) \right) \left(
\Gamma\left(B^0 \longrightarrow J/\psi \, K_s\right) +
\Gamma\left(\overline{B}^0 \longrightarrow J/\psi \, K_s\right)
\right)}    \nn  \\ 
&&\nn \\
&=& \sin 2\beta 
\eea
\bea
a_{\pi^+ \pi^-} & = & 
 -\frac{\Gamma\left(B^0 \longrightarrow \pi^+ \pi^-\right)
 - \Gamma\left(\overline{B}^0 \longrightarrow \pi^+ \pi^-\right)}{
 \left(\sin(\Delta M t ) \right) \left(
\Gamma\left(B^0 \longrightarrow \pi^+ \pi^-\right) +
\Gamma\left(\overline{B}^0 \longrightarrow \pi^+ \pi^-\right)
\right)}   \nn  \\
&&\nn \\
& = & \sin 2\alpha 
\eea
where we have adopted the standard definitions of the angles $\alpha$,
$\beta$ and $\gamma$ of the unitarity triangle
\bea
\label{unitangles}
\alpha \equiv \arg \left( -\frac{V_{td}V_{tb}^{*}}{V_{ud}V_{ub}^{*}}\right)
 \nn \\ 
\beta \equiv \arg \left( -\frac{V_{cd}V_{cb}^{*}}{V_{td}V_{tb}^{*}}\right)
\\ 
\gamma \equiv \arg \left( -\frac{V_{ud}V_{ub}^{*}}{V_{cd}V_{cb}^{*}}\right) 
\nn
\eea

In this letter, we will analyse how the presence of new physics
can be detected, once the asymmetries $a_{J\psi K_S}$, $a_{\pi^+ \pi^-}$ 
are measured. A good part of our analysis is applicable to a large class of 
models, although we pay special attention to the detection of 
flavour-changing neutral currents (FCNC) as well as deviations from
$3 \times 3 $ unitarity of the CKM matrix. In our analysis we will make 
the following assumptions:
\begin{itemize}
\item We will assume that the quark decay amplitudes $\overline{b}
\ra \bar{c}
c \bar{s}$, $\bar{b} \ra \bar{u} u \bar{d}$, as well as the semileptonic
$b$ decays are dominated by the SM tree-level diagrams. This is a reasonable
hypothesis, which is satisfied in most of the known extensions of the SM.

\item We will allow for the possibility of having new contributions to 
$B$--$\bar{B}$ mixing as well as deviations from $3 \times 3$ unitarity of 
the CKM matrix. 
\end{itemize}

We will define two new angles, $\alpha_{SM}$, $\beta_{SM}$, which have 
the interesting feature of being readily obtained from the measured values
of $a_{J/\psi K_S}$, $a_{\pi^+ \pi^-}$, independently of the presence of
new physics. We then indicate  how the values of $\alpha_{SM}$, 
$\beta_{SM}$ can be used to detect in a quantitative way the presence 
of new physics. This part of the analysis uses as experimental input
only the values of $a_{J/\psi K_S}$, $a_{\pi^+ \pi^-}$ and the ratio
$R_u\equiv |V_{ud}V_{ub}^*|/|V_{cd}V_{cb}^*|$. Using then the experimental 
value of the $B^{0}$--$\overline{B}^{0}$ mixing parameter $x_d$, we will show 
how deviations of $3 \times 3$ unitarity can be established by
full reconstruction of a unitarity quadrangle in the context of models 
extended with one isosinglet vector-like quark of the down type (VLdQ)
\cite{vlq2}.
We will show that CP asymmetries in B decays provide a very sensitive probe
on deviations from $3\times3$ unitarity, measured by the parameter $Z_{bd}$,
defined by
\beq
Z_{bd}=V_{ud}V_{ub}^{*}+V_{cd}V_{cb}^{*}+V_{td}V_{tb}^{*}.  \label{ZBD}
\eeq
We will give the minimum value of $Z_{bd}$ that can be detected at
$B$-factories, taking into account the expected experimental errors.

\section{Model-independent analysis}

It is clear from Eq. (\ref{unitangles}) that the angles $\alpha$, $\beta$,
$\gamma$ satisfy, by definition, the relation $\alpha +\beta +\gamma =
\arg(-1) = \pi $. This relation obviously holds in any model, and one can 
write $\gamma =\pi-(\alpha +\beta )$. We will allow for the possibility
of having new physics in the $B$--$\overline{B}$ mixing, which we will parametrize 
as 
\beq
M_{12}=M_{12}^{(SM)}\Delta _{bd}^{*}  \label{mixing}
\eeq
where $M_{12}^{(SM)}$ is the standard box contribution and $\Delta _{bd}$
is a complex number that parametrizes the new physics. The CP asymmetries
are then given by 
\beq
\begin{array}{l}
a_{J/\psi K_{s}}=\sin \left( 2\beta -\arg \Delta _{bd}\right) \\ 
a_{\pi ^{+}\pi ^{-}}=\sin \left( 2\alpha +\arg \Delta _{bd}\right).
\end{array}
\label{asyBSM}
\eeq
From this equation it is clear that $\alpha +\beta $ can be extracted in a 
model-independent way, and one has 
\beq
\pi -\gamma= \left( \alpha +\beta \right) =\frac{1}{2}\left[ 
\arcsin (a_{J/\Psi,K_{s}})+\arcsin (a_{\pi ^{+}\pi ^{-}})\right].  
\label{alp+bet}
\eeq

At this stage, it is useful to introduce the two angles $\alpha_{SM}$ and 
$\beta_{SM}$, defined by (see Fig. 1)
\bea
\alpha _{SM}=\arg \left[ \frac{\left(
V_{ud}V_{ub}^{*}+V_{cd}V_{cb}^{*}\right) }{V_{ud}V_{ub}^{*}}\right] \nn \\
\beta _{SM}=\arg \left[ \frac{V_{cd}V_{cb}^{*}}{\left(
V_{ud}V_{ub}^{*}+V_{cd}V_{cb}^{*}\right) }\right].
\label{asySM}
\eea

In models  that respect 3$\times$ 
3 unitarity, and in particular where $Z_{bd}=0$,
one obviously has $\alpha=\alpha _{SM}$ and $\beta =\beta _{SM}$, but this 
will not be the case in models where $Z_{bd}\neq 0$. The advantage of the new 
angles $\alpha_{SM}$, $\beta_{SM}$ results from the fact that they can be 
readily obtained from the measurements of $a_{J/\psi K_S}$, $a_{\pi^+ \pi^-}$
together with the ratio $R_u$. Indeed from Eq. (\ref{asySM}), one has, 
\bea
\label{alphaSM}
\alpha _{SM}=\arctan \left[ \frac{\sin \gamma }{R_{u}+\cos (\pi -\gamma )}%
\right] \nn \\ 
\beta _{SM}=\arctan \left[ \frac{R_{u}\sin \gamma }{1+R_{u}\cos (\pi -\gamma
)}\right], 
\eea
where $R_u=|V_{ud}V_{ub}^*|/|V_{cd}V_{cb}^*|$ and $\gamma$ is obtained from 
Eq. (\ref{alp+bet}). It should be emphasized that even if new physics is 
present in $B$--$\overline{B}$ mixing and/or 
there are deviations from unitarity,
$\alpha_{SM}$, $\beta_{SM}$ are obtained in a model-independent way from
Eq. (\ref{alphaSM}).  

We are
 specially interested in detecting any deviation of the measured values 
of the asymmetries $a_{J/\psi K_s}$, $a_{\pi^+ \pi^-}$ from the predictions 
of the standard model. These deviations can be defined as
\bea
\Delta_{J/\psi K_S} \equiv (a_{J/\psi K_S})_{measured} - (a_{J/\psi K_S})_{SM}
\nn \\
\Delta_{\pi^+ \pi^-} \equiv (a_{\pi^+ \pi^-})_{measured} - (a_{\pi^+ \pi^-})_{
SM},
\label{deltaSM}
\eea
where $(a_{J/\psi K_S})_{SM}$, $(a_{\pi^+ \pi^-})_{SM}$ are the predicted 
values of the asymmetries in the SM, namely:
\beq
\begin{array}{l}
(a_{J/\Psi ,K_{s}})_{SM}=\sin 2\beta _{SM} \\ 
(a_{\pi ^{+}\pi ^{-}})_{SM}=\sin 2\alpha _{SM},
\end{array}
\label{testNP}
\eeq

Since $\alpha_{SM}$, $\beta_{SM}$ can be evaluated from Eq. (\ref{alphaSM}),
one can obtain $\Delta_{J/\psi K_S}$, $\Delta_{\pi^+ \pi^-}$ having as 
experimental input only the experimental values of $a_{J/\psi K_S}$, 
$a_{\pi^+ \pi^-}$. Non-vanishing values of $\Delta_{J/\psi K_S}$, 
$\Delta_{\pi^+ \pi^-}$ indicate in a quantitative way the presence of
new physics.

For the analysis that follows, it is useful to define $\delta\equiv
\beta -\beta_{SM}$, which implies $\alpha =\alpha _{SM}-\delta $ (see Fig.
1). It is clear that the combination $2\delta -\arg \Delta _{bd}$ can
be readily evaluated from the previous analysis and one has
\bea
(2 \delta - \arg \Delta_{bd})= \arcsin (a_{J/\psi K_S}) - 2\beta_{SM}.
\label{dosdelta}
\eea
Notice that a deviation from zero in Eq. (\ref{dosdelta}) would translate in
a corresponding non-zero value in Eq. (\ref{deltaSM}). Therefore 
$2\delta- \arg\Delta_{bd}$ can also be used as a measure of the presence 
of new physics in CP asymmetries.

An additional piece of information that can be extracted from the previous
analysis is the side opposite to the angle $\gamma $ in the triangle with
angles $(\alpha _{SM},\beta _{SM},\gamma )$: 
\beq
L_{uc}\equiv |V_{ud}V_{ub}^{*}+V_{cd}V_{cb}^{*}|  \label{LUC}.
\eeq
With the experimental knowledge of $|V_{ud}V_{ub}^*|$, $|V_{cd}V_{cb}^{*}|$
as well as the value of $\gamma$ extracted from Eq. (\ref{alp+bet}), one
readily obtains $L_{uc}$.

So far, we have not used in our analysis the experimental value of
the $B^{0}$--$\overline{B}^{0}$ mixing parameter $x_{d}$. In the SM, 
$L_{uc}= |V_{td}V_{tb}^{*}|$; therefore, a second test of the SM comes 
from the comparison of $|L_{uc}|$ with the value of $|V_{td}V_{tb}^*|$
extracted from the value of $x_d$, in the framework of the SM.
If the equation
\beq
\begin{array}{l}
x_{d}=C_{d}  L_{uc}^{2} \\ 
\nn \\
C_{d}=\frac{G_{F}^{2}\eta_{B}M_{B_{d}}}{6\pi ^{2}\Gamma _{B_{d}}}%
(B_{B_{d}}F_{B_{d}}^{2})M_{W}^{2}\mid \overline{E}(x_{t} \mid)
\end{array}
\label{testluc}
\eeq
is not fulfilled, this will be a clear indication of the presence of new
physics beyond the SM. We have used standard notation for the parameters
entering  $C_d$, and their experimental values can be found in
Ref. \cite{buras}. In a general model, with a new contribution to
$B_d$--$\overline{B}_d$ mixing, one has
\beq
x_{d}=C_{d}\left| \Delta _{bd}\right| \left| V_{td}V_{tb}^{*}\right| ^{2}.
\label{XDgeneral}
\eeq

At this stage, it is worth recalling all the information we have 
about the unitarity quadrangle. From $\left| V_{ud}V_{ub}^{*}\right| $,
$\left|V_{cd}V_{cb}^{*}\right| $, $a_{J/\Psi ,K_{s}}$ and 
$a_{\pi ^{+}\pi ^{-}}$ we
have fully reconstructed the $(\alpha _{SM},\beta _{SM},\gamma )$-triangle.
The parameter $2\delta -\arg \Delta _{bd}$ is also obtained from Eq. 
(\ref{dosdelta}), while $x_{d}$ gives us the value of $|\Delta_{bd}|
|V_{td}V_{tb}^*|^2$. From Fig. 1 it is clear that, in order to obtain the 
full quadrangle, we need to reconstruct the triangle with sides $L_{uc}$,
$|Z_{bd}|$ and $|V_{td}V_{tb}^{*}|$.
Next we will indicate how the full quadrangle can be reconstructed 
in the specific case where the new contributions to $B_d$--$\overline{B}_d$
mixing are due to FCNC arising in the context of a model where the SM
is extended, through the addition of an isosinglet 
VLdQ. In this case $\Delta_{bd}$ is given by 
\bea
\label{deltabd}
\Delta _{bd}=1+are^{i\phi }-br^{2}e^{2i\phi }  \nn \\ 
re^{i\phi }=\frac{Z_{bd}}{V_{td}V_{tb}^{*}} \nn \\ 
a=\frac{4\overline{C}(x_{t})}{\overline{E}(x_{t})} \\ 
b=\frac{4\pi \sin ^{2}\theta _{W}}{\alpha \overline{E}(x_{t})}, \nn
\eea
where $\overline{E}(x_{t})$ and $\overline{C}(x_{t})$ are the well-known
Inami and Lim functions \cite{inami}: 
\bea
\overline{E}(x)=\frac{-4x+11x^{2}-x^{3}}{4(1-x)^{2}}+\frac{3x^{3}\ln x}{%
2(1-x)^{3}} \nn \\ 
\overline{C}(x_{t})=\frac{x}{4}\left[ \frac{4-x}{1-x}+\frac{3x\ln x}{%
(1-x)^{2}}\right]
\label{inamilim}
\eea
and $x_{t}=\left( m_{t}/M_{W}\right) ^{2}$. The last term in $\Delta _{bd}$, 
with an $r^2$ dependence, arises from the well-known
$Z$-flavour-changing tree-graph contribution. The second term, with a
linear dependence in $r$, comes from a one-loop $Z$-vertex 
correction, as recently pointed out in Ref. 
\cite{gabyquico}. 
From Eq. (\ref{deltabd}) and Fig. 1, it is evident that 
we are led, from the knowledge of
$|\Delta _{bd}| |V_{td}V_{tb}^*|^2$, $2\delta -\arg \Delta _{bd}$ and 
$L_{uc}$, to three equations  with three unknowns $r$, $\phi$, 
$ |V_{td}V_{tb}^*|$. Note that these last three variables completely
fix the upper triangle, therefore $\delta$ and $L_{uc}$ can be written 
in terms of $r$, $\phi$ and $|V_{td}V_{tb}^*|$. We can thus reconstruct 
the triangle of sides
$L_{uc}$, $|V_{td}V_{tb}^*|$, $|Z_{bd}|$, completing 
in this way the reconstruction 
of the unitarity quadrangle.

It should be pointed out that discrete ambiguities may
occur in the extraction of the value of the various angles from the 
knowledge of the asymmetries. A detailed discussion of how to overcome 
these ambiguities through additional measurements can be found in
Ref. \cite{quinn}. Throughout this paper, we will assume 
that these ambiguities can be solved  by using additional information.
We also assume that in the case of $\pi^+ \pi^-$, possible complications,
which may arise due to penguin contributions to the decay amplitudes, can 
be dealt with by using the analysis proposed in Ref. \cite{gronau}.

\section{Quadrangle reconstructions}

In VLdQ models one can choose, without loss of generality, a weak basis
where the up quark mass matrix is diagonal. The mixing is then described
by the 4$\times$4 unitary matrix $V$, 
which diagonalizes the down quark mass matrix. 
$V_{CKM}$ is the upper 3$\times$4
submatrix of $V$ and the elements of the fourth row fix the FCNC couplings
of the $Z$-boson, $Z_{qq^{\prime }}=-V_{4q^{\prime }}V_{4q}^{*}$. The actual
experimental bounds on $V$ coming from tree-level processes are 
\cite{buras}: 
\beq
\left[ 
\begin{array}{llll}
0.973-0.975 & 0.2187-0.223 & 0.0024-0.0040 & - \\ 
0.208-0.240 & 0.83-1 & 0.037-0.043 & - \\ 
- & - & - & - \\ 
- & - & - & -
\end{array}
\right]  \label{Decaybounds}
\eeq
From rare decays such as $K^{+}\ra \pi ^{+}\nu \overline{\nu }$ and $%
b\ra Xl^{+}l^{-}$ one obtains \cite{bounds}: 
\beq
\begin{array}{l}
\left| Z_{sd}\right| \leq 4.8\times 10^{-5} \\ 
\left| Z_{bd}\right| ,\left| Z_{bs}\right| \leq 1.9\times 10^{-3}
\end{array}
\label{rarebounds}
\eeq
and finally, from $x_{d}$: 
\beq
6.9\times 10^{-3}\leq \left| V_{td}V_{tb}^{*}\right| \left| \Delta
_{bd}\right| ^{1/2}\leq 11.3\times 10^{-3}  \label{xdbound}
\eeq
where the hadronic uncertainties have been included. 

The purpose of our analysis is to show through specific examples how with the
knowledge of the experimental values of the CP asymmetries $a_{J/\psi K_S}$,
$a_{\pi^+ \pi^-}$, 
and with the experimental errors expected in $B$-factories, one
may be able to detect new physics beyond the SM. With the assumption that new 
physics arises from the VLdQ model, we will show that one can fully 
reconstruct the unitarity quadrangles. In our analysis we have adopted the
following strategy. We have made a scan of $4\times 4$ unitary matrices,
using one of the standard parametrizations \cite{quicochau}, which on
the one hand satisfy all the constraints in Eqs. (\ref{Decaybounds}),
(\ref{rarebounds}), (\ref{xdbound}) and, on the other hand, lead to 
predictions to $(a_{J/\psi K_S}, a_{\pi^+ \pi^-})$ that differ significantly
from the predictions of the SM. We have classified the solutions in two groups 
with the following features.
\begin{itemize}
\item \underline{Case I}. Relatively large value of the parameter $r$
(e.g. $r \approx 0.2$), leading to a large value of $\arg\Delta_{bd}$,
while the deviations from $3\times 3$ unitarity, entering in the
asymmetries through $\delta$, remain relatively small. Then the effects
of new physics in the asymmetries are clearly dominated by the mixing. 
\item \underline{Case II}. Small values of $r$ (e.g. $r\approx 0.05$)
and new physics both in the mixing, $\arg\Delta_{bd}$, and in the quadrangle
$\delta$.
\end{itemize}
 
For definiteness, we will consider two examples of unitary matrices belonging 
to each one of the cases. We will then consider two realistic situations
where one knows $(a_{J/\psi K_S}, a_{\pi^+ \pi^-})$, 
$\left| V_{ud}V_{ub}^{*}\right| $, $\left|V_{cd}V_{cb}^{*}\right| $,
$x_d$, only within some experimental errors. The central values of 
$a_{J/\psi K_S}$, $a_{\pi^+ \pi^-}$ are chosen as the values implied by
the above-mentioned unitary matrices, belonging to cases I and II. We
will show that in each of these two cases, one would be able to 
establish the existence of new physics (i.e. $r \neq 0$) and find the
allowed ranges of $r$ and $\phi$.

Let us consider the following explicit examples:
\begin{itemize}
\item  Case I: $r \simeq 0.185$, new physics in the mixing
\end{itemize}

\beq
\left| V\right| =\left[ 
\begin{array}{cccc}
0.\,97496 & 0.\,2223 & 3.\,9999\times 10^{-3} & 0.00\,492 \\ 
0.\,22229 & 0.\,97423 & 3.\,7998\times 10^{-2} & 4.\,9666\times 10^{-3} \\ 
4.\,8288\times 10^{-3} & 3.\,7284\times 10^{-2} & 0.\,97745 & 0.\,20782 \\ 
4.\,1991\times 10^{-3} & 8.\,8125\times 10^{-3} & 0.\,2077 & 0.\,97814
\end{array}
\right]  \label{modV1}
\eeq

\beq
\arg (V)=\left[ 
\begin{array}{cccc}
0 & 0 & -0.\,41888 & 2.\,0944 \\ 
-3.\,1414 & -1.\,1839\times 10^{-5} & 1.\,9116\times 10^{-6} & 1.\,6755 \\ 
2.\,6719 & 4.\,1595\times 10^{-2} & 3.\,1416 & 0 \\ 
-2.\,4556 & -0.\,68252 & -3.\,1407 & 3.\,1416
\end{array}
\right]  \label{phaseV1}
\eeq

The corresponding unitarity quadrangle is represented in Fig. 2.

In this case, one has a relatively large value of $r$( $r= 0.\,18478$), and
there is new physics in the mixing corresponding to  $\arg
(\Delta _{bd})=2.\,0659$ versus $\delta\equiv \beta-\beta_{SM} =0.156$.
There are clearly detectable effects in  $a_{J/\psi K_S}$ and 
$a_{\pi^+ \pi^-}$, as can be seen by comparing to the values of 
$\sin 2\beta_{SM}$ and $\sin 2 \alpha_{SM}$:
\bea
&a_{J/\psi K_S} =-0.\,90274 & a_{\pi^+ \pi^- }=0.\,28481 \nn \\
&\sin (2\beta _{SM})=0.\,58736 &\sin (2\alpha _{SM})=-0.\,99442
\label{versus}
\eea

\begin{itemize}
\item  Case II:  $r \simeq 5\times 10^{-2}$ , new physics in the quadrangle 
and the mixing 
\beq
\left| V\right| =\left[ 
\begin{array}{cccc}
0.\,97496 & 0.\,2223 & 3.\,9999\times 10^{-3} & 0.00\,492 \\ 
0.\,22197 & 0.\,97422 & 3.\,9989\times 10^{-2} & 4.\,9666\times 10^{-3} \\ 
0.0\,1316 & 3.\,7863\times 10^{-2} & 0.\,9773 & 0.\,20801 \\ 
3.\,1224\times 10^{-3} & 6.\,8898\times 10^{-3} & 0.\,20799 & 0.\,9781
\end{array}
\right]  \label{modV2}
\eeq
\end{itemize}

\beq
\arg (V)=\left[ 
\begin{array}{cccc}
0 & 0 & -2.\,9322 & -2.\,0944 \\ 
-3.\,1413 & -1.\,2891\times 10^{-5} & 2.\,4309\times 10^{-6} & -0.\,83776 \\ 
3.\,0284 & 2.\,9735\times 10^{-2} & 3.\,1416 & 3.\,1416 \\ 
1.\,7074 & 2.\,4479 & 6.\,2532\times 10^{-4} & 3.\,1416
\end{array}
\right]  \label{phaseV2}
\eeq

The corresponding unitarity quadrangle is represented in Fig. 3.

In this case one has a rather small value of $r$ 
( $r=5.\,0494\times 10^{-2}$), while $\arg (\Delta_{bd})=-6.\,8996\times 
10^{-2}$ and $\delta =4.\,9506\times 10^{-2}$. There are also detectable 
effects of new physics in the CP asymmetries, since one has:
\bea
&a_{J/\psi K_S}=0.\,29147 & a_{\pi^+ \pi^- }=0.\,1233\nn \\
&\sin (2\beta _{SM})=0.\,12741&\sin (2\alpha_{SM})=0.\,2875
\label{versus2}
\eea
These two examples are the starting point of our analysis of realistic 
situations where the input data are only known within some experimental 
errors. At this stage it is worth indicating how the unitarity quadrangle
can be reconstructed, taking as input data $a_{J/\psi K_S}$, 
$a_{\pi^+ \pi^- }$, 
$\left| V_{ud}V_{ub}^{*}\right| $, 
$\left|V_{cd}V_{cb}^{*}\right| $ and $x_d$. We will use the following
relations: 
\beq
L_{uc} =\left| V_{cd}V_{cb}^{*}\right| \sqrt{1+R_{u}^{2}\mp 
\sqrt{2}R_{u}\sqrt{1+\left[ \pm \sqrt{\left( 1-a_{J/\psi K_S}^{2}\right) \left(
1-a_{\pi^+ \pi^-}^{2}\right) }-a_{J/\psi K_S}a_{\pi^+ \pi^-}\right] }}  \label{reconsfor1}
\eeq
\beq
2\delta -\arg (\Delta _{bd})=2\arg \left\{ 1+R_{u}e^{-\frac{i}{2}\left[
\arcsin \left( a_{\pi^+ \pi^-}\right) +\arcsin \left( a_{J/\psi K_S}\right) \right]
}\right\} +\arcsin \left( a_{J/\psi K_S}\right)  \label{reconsfor2}
\eeq
\beq
e^{i\left( 2\delta -\arg \left( \Delta _{bd}\right) \right) }=\frac{%
(1-re^{i\phi )^{2}}\left( 1+are^{i\phi }-br^{2}e^{2i\phi }\right) ^{*}}{%
\left| 1-re^{i\phi }\right| ^{2}\left| 1+are^{i\phi }-br^{2}e^{2i\phi
}\right| }  \label{reconsfor3}
\eeq
\beq
\frac{x_{d}}{C_{d} L_{uc}^2}=\frac{\left| 1+a re^{i\phi
}-br^{2}e^{2i\phi }\right| }{\left| 1-re^{i\phi }\right| ^{2}},
\label{reconsfor4}
\eeq
where we have expressed $|V_{td} V_{tb}^*|$
in terms of $r$, $\phi$ and $L_{uc}$ using, 
\bea
|V_{td} V_{tb}^*|= \frac{ L_{uc}}{|1 - r e^{i\phi}|}.
\label{lambdat}
\eea 
It is clear that using Eqs. (\ref{reconsfor1}), (\ref{reconsfor2})
and the measured values of $a_{J/\psi K_S}$, $a_{\pi^+ \pi^- }$, 
$\left| V_{cd}V_{cb}^{*}\right| $ and $\left|V_{ud}V_{ub}^{*}\right|$ 
one can obtain $ L_{uc}$ and $2\delta -\arg (\Delta _{bd})$.
One can then use Eqs. (\ref{reconsfor3}), (\ref{reconsfor4}) and
the measured value of $x_{d}$ to obtain $r$ and $\phi $. Therefore we can
fully reconstruct the quadrangle and obtain the important quantity $%
re^{i\phi }=\frac{Z_{bd}}{V_{td}V_{tb}^{*}}$, which is a measure of the
deviation of 3$\times$ 3 unitarity in the framework of the VLdQ model.

Next, we will assume that the measurement of $a_{J/\psi K_S}$, 
$a_{\pi^+ \pi^- }$ gives the central values corresponding to Case I, with
the following experimental errors:
\bea
a_{J/\psi K_S}=-0.90\pm 0.08\nn \\
a_{\pi^+ \pi^- }=0.28\pm 0.08
\label{expI}
\eea
where the errors probably are pessimistic for $a_{J/\psi K_S}$ and 
optimistic for $a_{\pi^+ \pi^- }$ in the case
of a $B$-factory. From these ``experimental'' data, together with $\left|
V_{cd}V_{cb}^{*}\right| $, $\left| V_{ud}V_{ub}^{*}\right| $ (including of
course their actual errors) and the experimental value of $x_{d}$ (with
the hadronic uncertainties included in $C_{d}$) we get in this case 
\beq
\begin{array}{l}
\cos \left( 2\delta -\arg (\Delta _{bd})\right) =0.06\pm 0.31 \\ 
\sin \left( 2\delta -\arg (\Delta _{bd})\right) =-0.998\pm 0.017 \\ 
0.90\leq \frac{x_{d}}{C_{d}\left| L_{uc}\right| ^{2}}\leq 5.13.
\end{array}
\label{bound1}
\eeq
The values given in Eq. (\ref{bound1}) can be plugged in  Eqs.
(\ref{reconsfor3}) and (\ref{reconsfor4})
in order to obtain $re^{i\phi }=x+iy$. Note that the value of the 
angle $2\delta -\arg(\Delta _{bd})$ 
would be quite far from its value in the SM, 
where it vanishes. Therefore, in this case the detection of new physics 
would be unambiguously
established. Note that in the derivation of the result of Eq. (\ref{bound1}),
we have assumed that the discrete ambiguities in Eqs. (\ref{reconsfor1}),
(\ref{reconsfor2}) can be solved by using
additional information, as pointed out in Ref. \cite{quinn}. The plot of
Eqs. (\ref{reconsfor3}) and (\ref{reconsfor4}), with the allowed bands
(Eq. (\ref{bound1})) is given in Fig. 4.

The region between the external circle and the two internal ones is allowed
by the $x_{d}$ ranges. The overlap of this sector with the allowed
regions for $\sin \left( 2\delta -\arg (\Delta _{bd})\right) $ and $\cos
\left( 2\delta -\arg (\Delta _{bd})\right) $ is the solution region we get
for $re^{i\phi }=x+iy$ (darker area). The input values in Case I are $x=-0.075$
and $y=-0.17$, so the region in the third quadrant of Fig. 4
corresponds to the inputs of Case I. To be sure that the other
solution in Fig. 4 is physical, it would be necessary to perform a very
fine scanning of the parameter space of the mixing matrix. We will not
address this question in this paper and we will concentrate on the
reconstructed solution that corresponds to the input parameters. Therefore we
conclude that roughly speaking we get a value of $r$ between $0.1$ and 
$0.2$, never compatible with the standard model, even if we enlarge the
error bars by several standard deviations. The two seagull-shaped regions
correspond to the $\cos \left( 2\delta -\arg \left( \Delta _{bd}\right)
\right) $ curves, the $\sin \left( 2\delta -\arg \left( \Delta _{bd}\right)
\right) $ curve serves to eliminate the wings in the second and fourth
quadrants. It is evident from equation (\ref{reconsfor3}) that the curves $%
\cos \left( 2\delta -\arg \left( \Delta _{bd}\right) \right) =cst$ will pass
through the points $\Delta _{bd}=0$ that correspond to the ``centre'' of the
small circles 
\bea
x&=&\frac{a}{2b}=9.\,92\times 10^{-3}\nn \\
y&=&\pm \sqrt{-\frac{1}{b}\left( 1+\frac{a^{2}}{2b}\right) }=\pm 
7.\,66\times 10^{-2}\\
r&=&0.0\,77\nn
\eea
It is precisely this value of $r$ that fixes the scale of $r$ lower bound
in Fig. 4. The upper bound of $r$ is fixed by the upper bound of $%
\frac{x_{d}}{C_{d}\left| L_{uc}\right| ^{2}}$ (the external circle);
therefore the larger $L_{uc}$ can be, the better upper bound 
for $r$. In this case, as it is evident from Fig. 2, $L_{uc}=\left(
6.3\pm 1.1\right) \times 10^{-3}$ is small. From this simple consideration,
we can expect to obtain in Case II a much better upper bound for $r$, since
from Fig. 3 one can see that $L_{uc}$ has a larger value.

Finally, let us consider Case II, where we assume the following values for
$a_{J/\psi K_S}$, $a_{\pi^+ \pi^-}$:
\bea
a_{J/\psi K_S}=0.29\pm 0.08, & a_{\pi^+ \pi^- }=0.12\pm 0.08,
\label{expII}
\eea
which lead to
\beq
\begin{array}{l}
\cos \left( 2\delta -\arg (\Delta _{bd})\right) =0.98\pm 0.01 \\ 
\sin \left( 2\delta -\arg (\Delta _{bd})\right) =0.19\pm 0.07 \\ 
0.274\leq \frac{x_{d}}{C_{d}\left| L_{uc}\right| ^{2}}\leq 1.10.
\end{array}
\label{bound2}
\eeq
Several comments are in order. The angle $2\delta -\arg (\Delta _{bd})$ is 
three standard deviations away from its standard model value of 0. This
is to be expected, since $a_{J/\psi K_S}$ is also three standard deviations
away from $\sin \left( 2\beta
_{SM}\right) $. The upper bound of $\frac{x_{d}}{C_{d} L_{uc}^{2}}$ is much 
smaller than in Case I as we have anticipated above. As we will
see these facts completely change the origin of the scales of the bounds in 
$r$ . The analog of Fig. 4 for the Case II is plotted in Fig. 5.

The input parameters in this second case are $x=0.012$ and $y=-0.049$,
therefore the solution
corresponds to the dark area in the
fourth qua\-drant. The so\-lu\-tions have chan\-ged 
qua\-drants because  $\sin \left( 
2\delta  \right.- $ $ \left. \arg (\Delta _{bd})\right) $ has 
changed sign. Clearly in this case
the upper bound on r ($r\lessim 0.065$) is fixed by the centre of the down
small circle defined by $\frac{x_{d}}{C_{d}\left| L_{uc}\right| ^{2}}=0.274$.
A reduction of the hadronic uncertainties in $C_{d}$ would lead to an increase
of the radius of this small circle, thus reducing the upper bound on $r$. 
The lower bound on $r$ ($r \gtrsim 0.04$) has to be taken with some caution,
since it is compatible with zero at less than three standard deviations.

At any rate, it is clear from the above analysis that the study of CP 
asymmetries at $B$-factories provides the possibility of putting 
stringent bounds on $\left| \frac{Z_{bd}}{V_{td}V_{tb}^{*}}\right| $,
of the order of 0.065 or even smaller. Therefore $B$-factories are quite 
competitive with respect to other experiments
looking directly for FCNC in $b\ra d$ transitions.

\section{Conclusions}

We have pointed out that the study of CP asymmetries in B decays provides an excellent 
tool for testing the SM and probing for new physics. In the framework 
of the SM, the CP
asymmetries $a_{J/\psi K_s}$, $a_{\pi^+\pi^-}$  measure the angles of the unitarity
triangle. If one wants to allow for the possiblity of having new physics in the 
$B$--$\overline{B}$ mixing
and/or deviations of 3$\times$3 unitarity of the CKM matrix, the 
analysis becomes 
more involved. We found it useful to introduce two new angles, 
$\alpha_{SM}$ and $\beta_{SM}$, whose important feature is the fact that 
they are readily obtained from the measurement of $a_{J/\psi K_s}$ , $a_{\pi^+\pi^-}$ and
the ratio $R_u$, even in the presence of new physics. The knowledge of 
$\alpha_{SM}$ and $\beta_{SM}$  enables us to
ckdetect in a quantitative way deviations from the SM predictions for the CP asymmetries.

Assuming that the new physics arises from deviations from 3$\times$3 unitarity,
 which in turn leads
to FCNC, we show how to reconstruct the unitarity quadrangle from input data. A study of 
specific examples shows that even taking into account the expected errors in the 
experimental values of $a_{J/\psi K_s}$, $a_{\pi^+\pi^-}$, it may be 
possible to detect the 
presence of the FCNC coupling $Z_{bd}$,  even for a rather small value of the ratio 
$\left|\frac{Z_{bd}} {V_{td}V_{tb}^{*}}\right|$.

\vspace{.5cm}

{\bf ACKNOWLEDGEMENTS}

We would like to thank J. Bernab\'eu and F. del Aguila
for useful comments and
discussions.
G.B. acknowledges the Spanish Ministry of
Foreign Affairs for a MUTIS fellowship, and O.V.
acknowledges the Generalitat Valenciana for a research
fellowship. G.C. Branco is grateful to the CERN Theory Division
for kind hospitality, and thanks Michael Gronau for various 
conversations.
 This work is partially supported by CICYT under 
grant AEN-96-1718 and IVEI under grant 038/96.

\clearpage

\begin{figure}
\begin{center}
\epsfxsize = 13cm
\epsffile{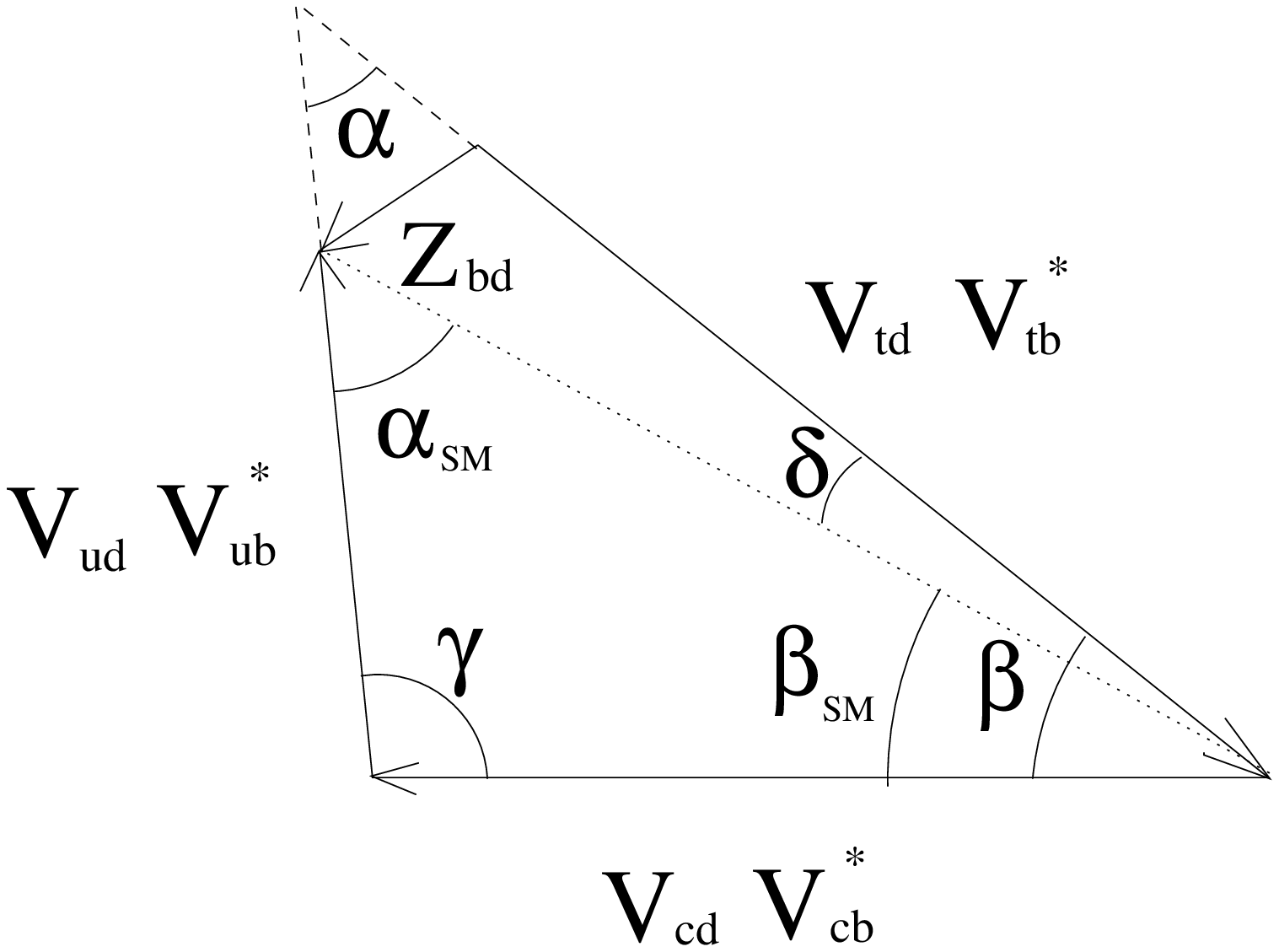}
\leavevmode
\end{center}
\caption{The unitarity quadrangle, with angles $\alpha$, $\beta$, $\gamma$,
$\alpha_{SM}$, $\beta_{SM}$. In the SM limit one has $\alpha_{SM} =
\alpha$, $\beta_{SM}=\beta$.}
\end{figure}

\clearpage

\begin{figure}
\begin{center}
\epsfxsize = 15cm
\epsffile{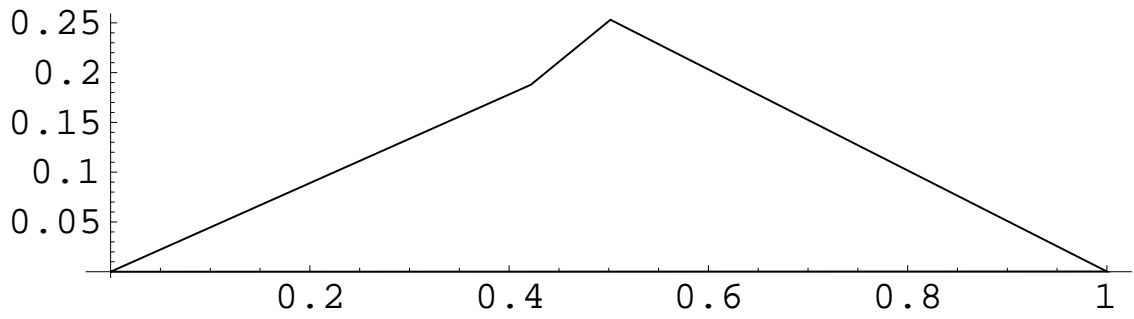}
\leavevmode
\end{center}
\caption{ The unitarity quadrangle for case I, corresponding to a 
relatively large contribution of new physics in the mixing,
$r=0.185$}
\end{figure}

\clearpage

\begin{figure}
\begin{center}
\epsfxsize = 13cm
\epsffile{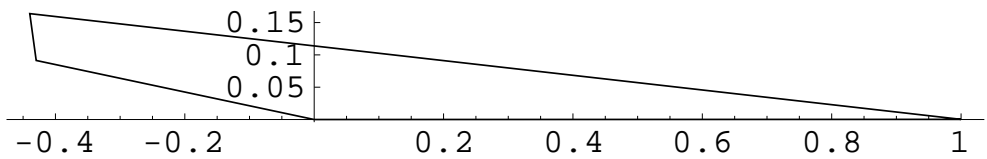}
\leavevmode
\end{center}
\caption{ The unitarity quadrangle for case II, corresponding to 
$ r = 5 \cdot 10^{-2} $}
\end{figure}

\pagebreak
\begin{figure}
\begin{center}
\epsfxsize = 13cm
\epsffile{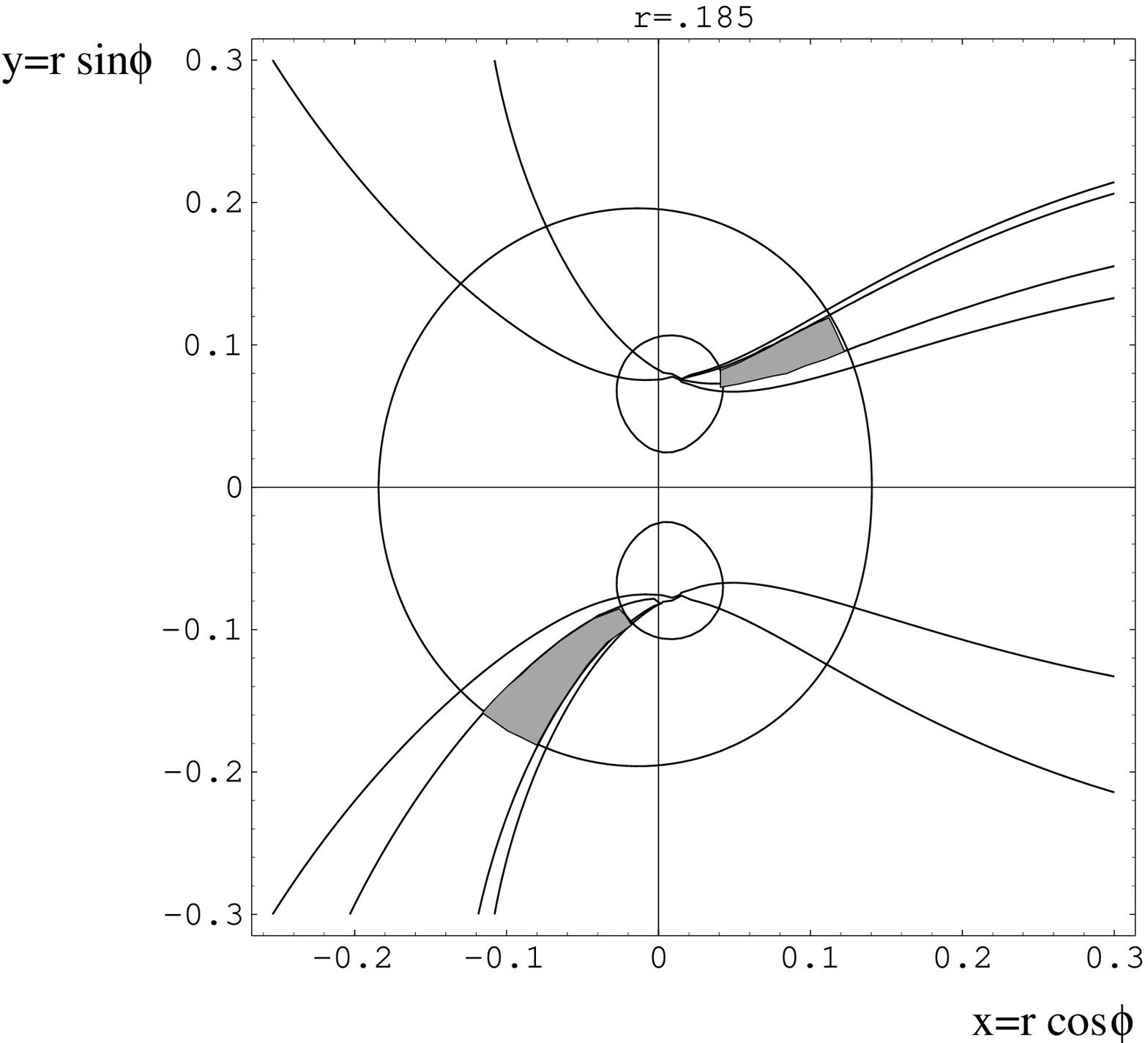}
\leavevmode
\end{center}
\caption{Reconstruction of $(r,\phi)$ from the measurement of the asymmetries
in Case I}
\end{figure}

\clearpage

\begin{figure}
\begin{center}
\epsfxsize = 13cm
\epsffile{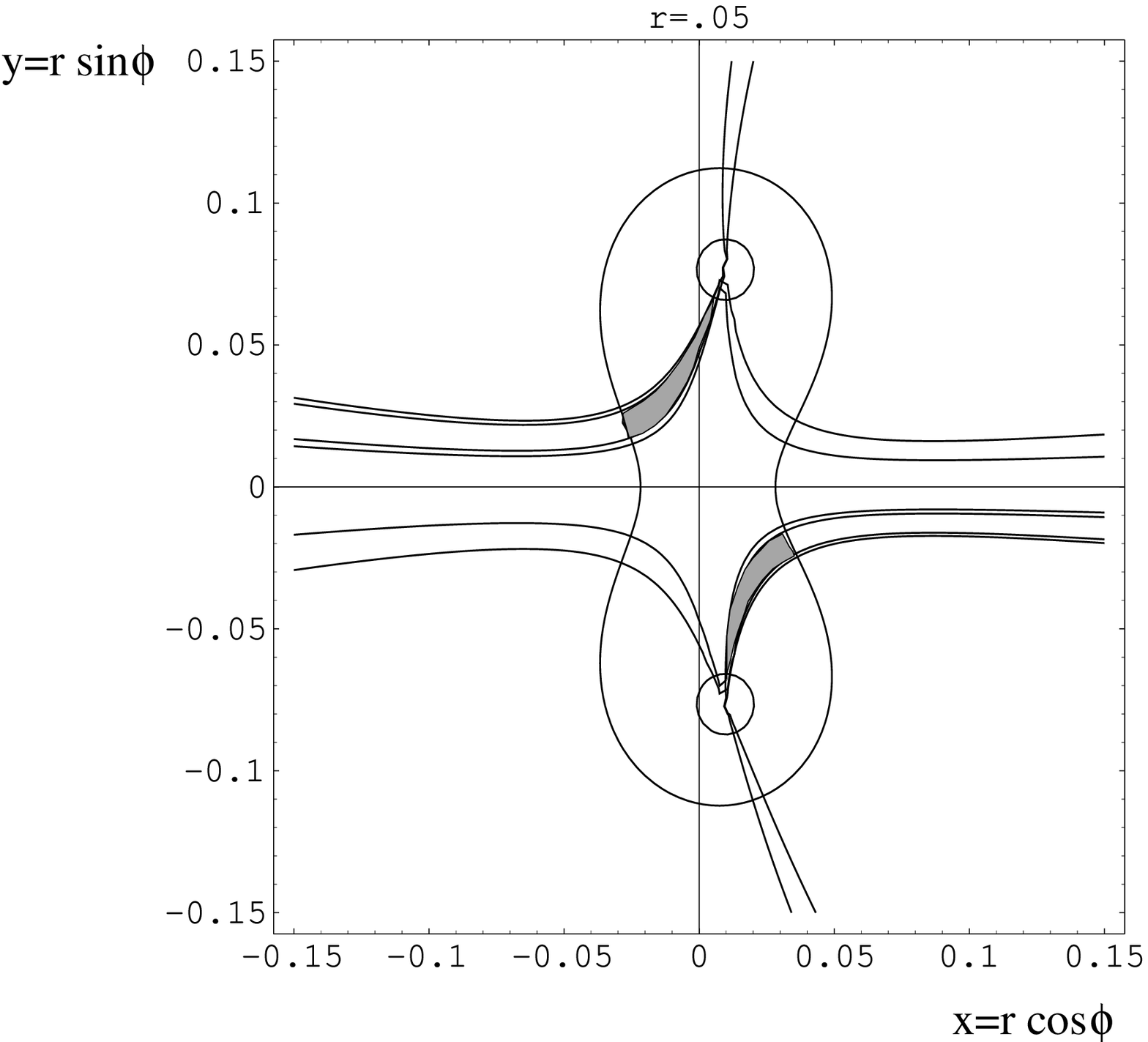}
\leavevmode
\end{center}
\caption{Reconstruction of $(r, \phi)$ from the measurement of the asymmetries
in Case II }
\end{figure}

\end{document}